\documentclass[twocolumn,english,aps,prd,reprint,superscriptaddress,floatfix,notitlepage,nobibnotes,nofootinbib,preprintnumbers]{revtex4-1}
\pdfoutput=1
\usepackage{lmodern}

\usepackage[T1]{fontenc}
\usepackage[latin9]{inputenc}
\usepackage{color}
\usepackage{comment}
\usepackage{babel}
\usepackage{amsmath}
\usepackage{amssymb}
\usepackage{graphicx}
\usepackage{stackengine}
\usepackage{esint}
\usepackage[unicode=true,pdfusetitle,
 bookmarks=true,bookmarksnumbered=false,bookmarksopen=false,
 breaklinks=false,pdfborder={0 0 1},backref=false,colorlinks=true]
 {hyperref}
\hypersetup{pdftitle={Entropy Bounds on Effective Field Theory from Rotating Dyonic Black Holes},
 pdfauthor={Clifford Cheung, Junyu Liu, and Grant N. Remmen},
 citecolor=black,linkcolor=black,urlcolor=black}
\usepackage{breakurl}
\usepackage[hang,flushmargin]{footmisc} 
\allowdisplaybreaks
\makeatletter

 \@ifundefined{textcolor}{}
 {%
   \definecolor{BLACK}{gray}{0}
   \definecolor{WHITE}{gray}{1}
   \definecolor{RED}{rgb}{1,0,0}
   \definecolor{GREEN}{rgb}{0,1,0}
   \definecolor{BLUE}{rgb}{0,0,1}
   \definecolor{CYAN}{cmyk}{1,0,0,0}
   \definecolor{MAGENTA}{cmyk}{0,1,0,0}
   \definecolor{YELLOW}{cmyk}{0,0,1,0}
 }

\definecolor{nicegreen}{rgb}{0.0, 0.5, 0.0}

\usepackage{babel}

\makeatletter
\def\simgt{\mathrel{\lower2.5pt\vbox{\lineskip=0pt\baselineskip=0pt
           \hbox{$>$}\hbox{$\sim$}}}}
\def\simlt{\mathrel{\lower2.5pt\vbox{\lineskip=0pt\baselineskip=0pt
           \hbox{$<$}\hbox{$\sim$}}}}
\makeatother

\newcommand{\be}{\begin{equation}}
\newcommand{\ee}{\end{equation}}
\newcommand{\bea}{\begin{eqnarray}}
\newcommand{\eea}{\end{eqnarray}}
\newcommand{\Ref}[1]{Ref.~\cite{#1}}
\newcommand{\Fig}[1]{Fig.~\ref{#1}}
\newcommand{\Tab}[1]{Table~\ref{#1}}
\newcommand{\Eq}[1]{Eq.~(\ref{#1})}

\newcommand{\App}[1]{App.~\ref{#1}}

\begin{document}

\pagestyle{plain}

\preprint{CALT-TH-2019-003}

\title{Entropy Bounds on Effective Field Theory from Rotating Dyonic Black Holes}

\author{Clifford Cheung}
\affiliation{Walter Burke Institute for Theoretical Physics\\ California Institute of Technology, Pasadena, CA 91125, USA}
\author{Junyu Liu}
\affiliation{Walter Burke Institute for Theoretical Physics\\ California Institute of Technology, Pasadena, CA 91125, USA}
\affiliation{Institute for Quantum Information and Matter\\ California Institute of Technology, Pasadena, CA 91125, USA}
\author{Grant N. Remmen}
\affiliation{Center for Theoretical Physics and Department of Physics\\
University of California, Berkeley, CA 94720, USA and\\
Lawrence Berkeley National Laboratory, Berkeley, CA 94720, USA}


\begin{abstract}
\noindent 
We derive new bounds on higher-dimension operator coefficients in four-dimensional Einstein-Maxwell theory.  Positivity of classically-generated corrections to the Wald entropy of thermodynamically stable, rotating dyonic black holes implies a multiparameter family of field basis invariant inequalities that exhibit electromagnetic duality and are satisfied by examples from field and string theory. These bounds imply that effective operators modify the extremality condition of large black holes so as to permit their decay to smaller ones, thus satisfying the weak gravity conjecture.

\end{abstract}
\maketitle
\interfootnotelinepenalty=10000
\let\thefootnote\relax\footnote{e-mail: \url{clifford.cheung@caltech.edu}, \url{jliu2@caltech.edu},\\
\phantom{\hspace{10.1mm}}\url{grant.remmen@berkeley.edu}}

\setcounter{footnote}{0}

\section{Introduction}
The swampland program~\cite{Vafa:2005ui,Ooguri:2006in,ArkaniHamed:2006dz} is founded on the intriguing possibility that an infrared (IR) consistent effective field theory (EFT) may nevertheless be incompatible with quantum gravitational ultraviolet (UV) completion.  An archetype of this approach is the weak gravity conjecture (WGC)~\cite{ArkaniHamed:2006dz}, which states that any Abelian gauge theory coupled consistently to gravity must contain a state whose charge exceeds its mass in Planck units.  Other notable results from the swampland program include the distance conjecture \cite{Ooguri:2006in}, and AdS \cite{Ooguri:2016pdq} and dS~\cite{Obied:2018sgi,Ooguri:2018wrx} conjectures.

The very notion of the swampland is predicated on a complete knowledge of the space of IR consistent EFTs.  After all, if a putative swampland theory is also pathological from IR considerations, then UV completion has little to do with its sickness.  For example, gauge anomalies are absent in grand unified theories, but they are also ruled out by low-energy reasoning.   In parallel with the swampland program, there has emerged a bottom-up approach to constraining EFTs using IR properties like unitarity, causality, and analyticity of scattering amplitudes~\cite{Adams:2006sv}.
These tools can be used to derive new positivity conditions on higher-dimension operator coefficients in EFTs~\cite{Adams:2006sv,Cheung:2014ega,Bellazzini:2015cra,Cheung:2016yqr,GB+,Bellazzini:2019xts}, in some cases with broad implications, e.g., the proof of the four-dimensional $a$-theorem~\cite{Komargodski:2011vj}. 

At the same time, the study of black hole thermodynamics has produced many elegant and powerful constraints on entropy with important implications for both the IR and UV~\cite{Bekenstein:1974ax,Bekenstein:1980jp,Bousso:1999xy,Goon:2016mil}. In this paper we bring together these three lines of inquiry---the swampland program, EFT bounds, and entropy inequalities---to use black hole entropy to constrain the landscape of consistent EFTs.

We consider rotating dyonic black holes in the EFT of gravitons and photons in four spacetime dimensions, with action $\int {\rm d}^4 x\sqrt{-g} (\overline {\cal L}+ \Delta {\cal L})$, where the Einstein-Maxwell (EM) term is $\overline {\cal L} = \frac{1}{2\kappa^2}R - \frac{1}{4}F_{\mu\nu}F^{\mu\nu}$ and
\be
\begin{aligned}
\Delta {\cal L} = &\phantom{{} + {}} c_1 R^2 + c_2 R_{\mu\nu}R^{\mu\nu} + c_3 R_{\mu\nu\rho\sigma}R^{\mu\nu\rho\sigma} \\
& +c_4 R F^2 +c_5 R^{\mu\nu} F_\mu^{\;\;\rho}F_{\nu\rho}+ c_6 R^{\mu\nu\rho\sigma}F_{\mu\nu}F_{\rho\sigma} \\
& +c_7 F_{\mu\nu} F^{\mu\nu} F_{\rho\sigma} F^{\rho\sigma} + c_8  F_{\mu\nu}F^{\nu\rho}F_{\rho\sigma}F^{\sigma\mu},
\end{aligned} \label{eq:DeltaL}
\ee
in the sign conventions of \Ref{Gravitation} and where $\kappa^2 = 8\pi G$.
This action encodes all leading-order parity-conserving interactions of gravitons and photons at low energies.  In the absence of charged currents, any operator with derivatives on $F_{\mu\nu}$ can be recast into those already in \Eq{eq:DeltaL} using the Bianchi identities~\cite{Deser:1974cz}. 

In \Ref{Cheung:2018cwt} it was proven that when the higher-dimension operators in \Eq{eq:DeltaL} are generated at tree level by quantum field theoretic dynamics, they induce a positive shift to the Wald entropy of a thermodynamically stable black hole at fixed mass and charge,
\be 
\Delta S > 0.\label{eq:DSbound}
\ee 
This bound applies for black holes sufficiently small so that the classical Wald entropy shift dominates over quantum corrections, which is always achievable within the regime of validity of a weakly coupled EFT \cite{Cheung:2018cwt}; similar power counting shows that Hawking radiation can be ignored. The physical origin of positivity follows from an intimate linkage between the shift in entropy and the variation of the Euclidean action away from its local minimum, which is positive for thermodynamically stable systems.  The extensive details can be found in \Ref{Cheung:2018cwt}.

As shown in \Ref{Kats:2006xp}, higher-derivative operators in the EFT modify the extremality condition for black holes.  In \Ref{Cheung:2018cwt} it was shown that  $\Delta S >0$ implies constraints on $c_i$ that {\it precisely} tip the extremality condition so that large electrically charged black holes are unstable to decay to smaller ones.  Consequently, the WGC is automatically satisfied since there exist states---namely, black holes---with charge-to-mass ratio greater than unity.    More generally,  \Ref{Cheung:2018cwt} showed that $\Delta S >0$ implies the WGC for any number of Abelian forces and in arbitrary spacetime dimension $\geq 4$.

In this paper we consider four-dimensional rotating dyonic black holes, generalizing the results of \Ref{Cheung:2018cwt} to obtain a family of constraints on the coefficients $c_i$, labeled by the angular momentum and electric and magnetic charge-to-mass ratios of the black hole.  As before, $\Delta S>0$ exactly guarantees that the extremality condition tips so that large black holes are unstable, thus establishing a dyonic rotating version of the WGC.

\section{Metric and Action}
To begin, we  compute the Wald entropy for a black hole of fixed electric charge $Q$, magnetic charge $\widetilde Q$, angular momentum $J$, and mass $M$ at leading order in the higher-dimension operator coefficients $c_i$ in \Eq{eq:DeltaL}.
For later convenience we introduce natural quantities
\be 
\begin{aligned} 
m &= \kappa^2 M/8\pi\\
q &= \kappa Q/4\sqrt{2}\pi\\
\widetilde q&= \kappa \widetilde Q/4\sqrt{2}\pi
\end{aligned}
\ee
 and rescaled coefficients
 \be 
 \begin{aligned}
 d_{1,2,3} &= \kappa^2 c_{1,2,3} \\
 d_{4,5,6} &= c_{4,5,6}\\
 d_{7,8} &= \kappa^{-2} c_{7,8}.
 \end{aligned}
 \ee 
 The $d_i$ all scale as  $1/\Lambda^2$, where $\Lambda$ is the scale of a weakly coupled UV completion.  

The unperturbed Kerr-Newman (KN) metric $\overline g_{\mu\nu}$ in Boyer-Lindquist coordinates is given by
\be
\begin{aligned}
{\rm d}s^{2}=&-\frac{\Delta}{\rho^{2}}({\rm d}t-a\sin^{2}\theta\,{\rm d}\phi)^{2}+\rho^{2}\left(\frac{{\rm d}r^{2}}{\Delta}+{\rm d}\theta^{2}\right)\\&+\frac{\sin^{2}\theta}{\rho^{2}}\left[\left(r^{2}+a^{2}\right){\rm d}\phi-a\,{\rm d}t\right]^{2},
\end{aligned}\label{eq:KNmetric}
\ee
where 
\be 
\begin{aligned}
\rho^2 &=r^2+a^2 \cos^2 \theta\\
a &= J/M = \kappa^2 J/8\pi m\\ 
\Delta &=r^2-2mr+a^2+q^2+\widetilde q^2.
\end{aligned}
\ee
It will be convenient to define 
\be 
\xi = \frac{\sqrt{m^2-(q^2+\widetilde q^2+a^2)}}{m},
\ee 
so that $\xi \rightarrow 0$ is the extremal limit.
The event horizon is located at $r$-coordinate $r_{\rm H} = m(1+\xi)$.
We also define parameters $\mu =(q^2 - \widetilde q^2)/(q^2 + \widetilde q^2)$ and $\nu ={a}/{r_{\rm H}}$ to characterize the charge and spin,
so $\mu = 1$ and $\mu= -1$ correspond to pure electric and pure magnetic black holes, respectively, $\nu = 0$ to nonrotating charged black holes, and $\nu = \sqrt{(1-\xi)/(1+\xi)}$ to uncharged rotating black holes.

Next, we employ a relation between entropy and the Euclidean action that was discussed in \Ref{Cheung:2018cwt} and recently generalized and proven in \Ref{Reall:2019sah} via a simple argument from basic thermodynamic identities:
\be
\Delta S = - \Delta I ,\label{eq:SI}
\ee
where $\Delta S$ is the entropy shift at fixed mass, charge, and angular momentum, and $\Delta I$ denotes the higher-dimension operator contributions to the action evaluated on the leading-order black hole solution, working in an ensemble with fixed inverse temperature $\beta$, angular velocity $\Omega$, and charge.

We will compute the entropy shift $\Delta S$ to ${\cal O}(d_i)$ using \Eq{eq:SI} for a rotating dynonic black hole.  Since KN is a stationary spacetime with Killing vector $\partial_t$, we have 
\be 
\Delta I = - \beta \int {\rm d}^3 x \sqrt{-g}\Delta {\cal L}|_{\rm KN},
\ee 
with the right-hand side evaluated on the Lorentzian KN black hole background at fixed $t$, where the integration is over all $(r,
\theta,\phi)$ outside the horizon. 

Note that the Gauss-Bonnet term ${\cal L}_{\rm GB} = R^2 - 4R_{\mu\nu}R^{\mu\nu}+R_{\mu\nu\rho\sigma}R^{\mu\nu\rho\sigma}$ corrects black hole entropy \cite{Clunan:2004tb} even though it is a total derivative in four dimensions.  This is not contradictory since the black hole horizon provides a boundary on which total derivatives  have support.  This is similar in spirit to what happens to $F_{\mu\nu} \widetilde F^{\mu\nu}$ in electrodynamics, which is a total derivative but comes into play in the background of a magnetic monopole.

Nevertheless, one can show that are no other total derivative or boundary operators that correct the entropy at the leading order of interest.  For example,  $\Box R$ and $\Box (F_{\mu\nu}F^{\mu\nu})$ do not affect the Wald entropy nor the equations of motion.  Meanwhile, higher-derivative boundary terms at asymptotic infinity, analogous to the Gibbons-Hawking-York term, are subleading in $1/r$ at large distances and can therefore be ignored \cite{Reall:2019sah,Cheung:2018cwt}.

\section{Black Hole Entropy and Consistency Checks}
Any physical constraint on the action must be invariant under change of field basis, e.g., the metric transformation $g_{\mu\nu} \rightarrow g_{\mu\nu} + r_1 R_{\mu\nu} + r_2 Rg_{\mu\nu} + r_3 F_{\mu\rho}F_\nu^{\;\;\rho} + r_4 F_{\rho\sigma}F^{\rho\sigma} g_{\mu\nu}$.  Consequently, a given coefficient $d_i$ is not individually meaningful because it mixes with others under a field redefinition.  
The four field-redefinition invariant linear combinations of coefficients are  $(d_0,d_3,d_6,d_9)$, where \cite{Cheung:2018cwt}
\be
\begin{aligned}
d_0 &= d_2 + 4d_3 + d_5 + d_6 + 4d_7 + 2d_8 \\
d_9 &= d_2 + 4d_3 + d_5 +2 d_6 + d_8.
\end{aligned}\label{eq:d0d9}
\ee
Note that \Eq{eq:SI} is automatically field-basis invariant because those combinations of higher-dimension operators that are removable by a field redefinition are proportional to equations of motion and thus vanish when evaluated on the leading-order solutions.
We find that $\Delta S$ is indeed field-basis invariant:
\begin{widetext}
\be
\begin{aligned}
\Delta S(\xi,\mu,\nu) =&\phantom{{}+{}}\frac{16\pi^2 (2\mu^2-1)(\nu^2-3)(3\nu^2-1)[1-\xi-\nu^2(1+\xi)]^2}{15\kappa^2\xi(1+\xi)(1+\nu^2)^5}(d_0+d_6-d_9) \\
& + \frac{\pi^2[1-\xi-\nu^2(1+\xi)]^2 [\nu(3+2\nu^2+3\nu^4)+3(\nu^2-1)(1+\nu^2)^2 \,{\rm arctan}\,\nu]}{2\kappa^2\xi(1+\xi)\nu^5 (1+\nu^2)}(d_0+d_6+d_9)\\
& + \frac{64\pi^2}{\kappa^2(1+\nu^2)} d_3 + \frac{32\pi^2 \mu [1-\xi-\nu^2(1+\xi)][\nu^2(3+4\xi)-1-4\xi]}{5\kappa^2 \xi(1+\xi)(1+\nu^2)^3}d_6 .
\end{aligned}\label{eq:DSKN}
\ee
\end{widetext}

Note that \Eq{eq:DSKN} is also invariant under the electromagnetic duality transformation, $q\leftrightarrow\widetilde q$ and $F_{\mu\nu }\leftrightarrow\widetilde F_{\mu\nu} = \epsilon_{\mu\nu\rho\sigma}F^{\rho\sigma}/2$.
However, individual operators in $\Delta {\cal L}$ are not invariant, since $F_{ab} \rightarrow \widetilde F_{ab}$ sends $d_i \rightarrow \widetilde d_i$, where 
\be
\!\widetilde d_i \!= \! (d_1,d_2,d_3,- d_4 - d_5/2 - d_6, d_5 + 4d_6,  - d_6,d_7,d_8),\!
\ee
 i.e., 
 \be 
\widetilde d_{0,3,6,9} = (d_0 + 2d_6, d_3, -d_6, d_9).
\ee
Applying this mapping to the entropy shift $\Delta S(\xi,\mu,\nu)$ should be equivalent to swapping $|q|$ and $|\widetilde q|$, sending $\mu$ to $-\mu$. Indeed, from \Eq{eq:DSKN} we find that
\be 
\left. \Delta S(\xi,\mu,\nu)\right|_{d_i \rightarrow \widetilde d_i} = \Delta S(\xi,-\mu,\nu),
\ee
so our family of bounds transforms consistently under electromagnetic duality.

While \Eq{eq:DSKN} carries a formal divergence at $\xi = 0$,  as shown in \Ref{Cheung:2018cwt} we can take the near-extremal, i.e., small-$\xi$, limit consistent with perturbative control.  Requiring that the near-extremal shifts in entropy and temperature be much smaller than their unperturbed values mandates $\xi \gg \kappa \Lambda$, but $\kappa \Lambda$ is parametrically small for a field-theoretic completion.  For later convenience we will sometimes use $\xi=0$ to indicate this limiting value.

As an independent check of our calculation, we have used perturbative techniques~\cite{Campanelli:1994sj,Kats:2006xp} applicable to spherically-symmetric black holes to solve for the metric itself at leading order in the $d_i$, making use of the invertibility of the Ricci tensor for spherical geometries; see \App{app} for details.  Inserting the metric into the Wald formula, we obtain $\Delta S$ for the $\nu=0$ case, and we find that it agrees with \Eq{eq:DSKN}. 
As a further consistency check, we have verified that the surface gravity at the horizon in the perturbed metric agrees with the thermodynamic temperature extracted from differentiating the Wald entropy with respect to the mass.

\section{EFT Bounds}
We now show how $\Delta S>0$ constrains the EFT.  As discussed earlier, this inequality holds for black holes that are thermodynamically stable.
For a general KN black hole, the requirement of positive heat capacity at constant $J$ is not enough to guarantee thermodynamic stability.  In our ensemble at fixed $\beta$ and $\Omega$, we must also require positive isothermal moment of inertia~\cite{Monteiro:2009tc}.
This enhanced stability condition restricts our consideration to
\be 
0<\xi < \frac{1-3\nu^2}{2(1+\nu^2)}.\label{eq:stable}
\ee
In the $\nu =0$ nonrotating case, this window reduces to $\xi \in (0,\tfrac{1}{2})$, i.e., $q^2 + \widetilde q^2  > 3m^2/4$. The stability condition in \Eq{eq:stable} cannot be satisfied in the neutral case, where $\nu = \sqrt{(1-\xi)/(1+\xi)}$.  Thus, we cannot use neutral Kerr black holes to place a bound because they are thermodynamically unstable. In the $\xi\rightarrow 0$ extremal limit, \Eq{eq:stable} reduces to $\nu \in [0,1/\sqrt{3})$, i.e., $q^2+\widetilde q^2 > 2a^2$. 

As discussed earlier, the entropy shift satisfies
\be
\Delta S(\xi,\mu,\nu)>0\label{eq:boundfull}
\ee
for all $(\xi,\nu)$ satisfying \Eq{eq:stable} and all $\mu\in[-1,1]$, which is the main result of this work.  This condition implies new bounds on the EFT of gravitons and photons mandated by the entropy of rotating dyonic black holes. 

\begin{figure*}[ht]
\begin{center}
\includegraphics[width=\textwidth]{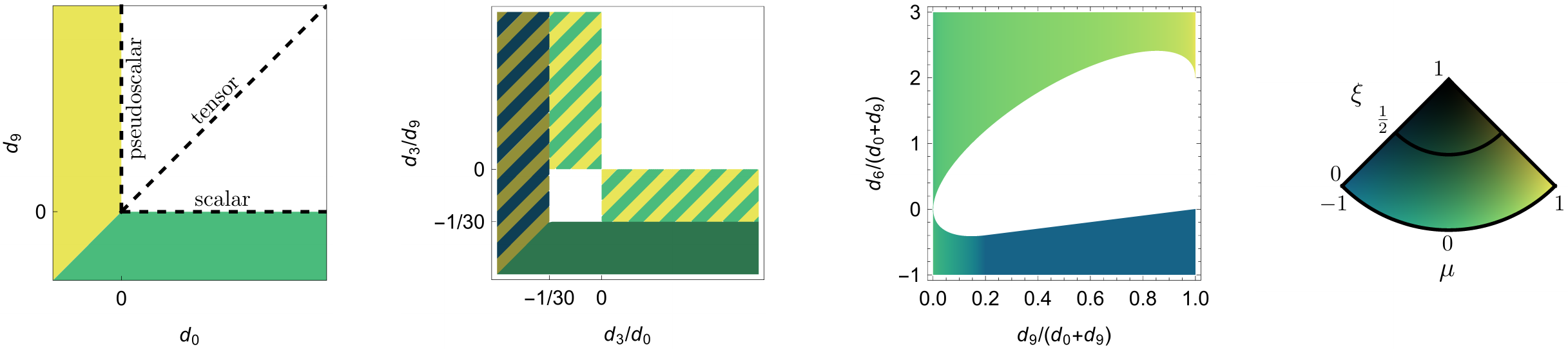}
\end{center}
\vspace{-6mm}
\caption{Entropy bounds on higher-dimension operator coefficients $d_i$.  The shaded regions are excluded since $\Delta S(\xi,\mu,0)<0$ for some choice of black hole (yellow for electric, blue for magnetic, with brighter hues indicating larger charge), irrespective of the values of operator coefficients not shown in a given panel. These constraints imply that $d_{0,9}$ are positive, $d_3$ is bounded from below, and $d_6$ is bounded by various combinations of $d_{0,9}$. For the scalar, pseudoscalar, and tensor UV completions in \Tab{tab:UV}, $d_3 = d_6 = 0$ and allowed values of $d_{0,9}$ are indicated by the black dashed lines in the leftmost panel.}
\label{fig:boundspace}
\end{figure*}

Numerical studies indicate that the parameter space excluded by \Eq{eq:boundfull} is already forbidden by the special case $\nu =0$, i.e., the bounds from rotating black holes are already accounted for by the nonrotating case.   See \Fig{fig:boundspace} for examples of excluded regions, which we now discuss.

Two immediate implications of \Eq{eq:boundfull} are that
\be 
d_0>0 \text{  and  } d_9>0,
\ee obtained by taking $\Delta S(0,1,0)>0$ and $\Delta S(0,0,0)>0$, respectively. Note that the four-dimensional bounds advocated via IR consistency arguments in \Ref{Cheung:2014ega} are a subset of \Eq{eq:boundfull}.
If only $d_{7,8}$ are nonzero, the bounds in this paper can be written simply as $2d_7+d_8>0$ and $d_8>0$, both of which are implied by analyticity of four-photon scattering \cite{Adams:2006sv,Cheung:2018cwt}.

\begin{table*}[t]
\begin{center}
\begin{tabular}{c| c| r| c } 
massive state & ${\cal L}_{\rm int}$&\multicolumn{1}{c|}{$d_{0,3,6,9}$} &$\Delta S(\xi,\mu,0)\times [5\kappa^2 \xi(1+\xi)/32\pi^2]$ \\ \hline
scalar & $\phi\left(a \kappa^{-1} R  + \kappa b F_{\mu\nu}F^{\mu\nu}\right)$ & $2b^2/m_\phi^2\times(1,0,0,0 )$ &  $2(1-\xi)^2 b^2 \mu^2/m_\phi^2$ \\ \hline
pseudoscalar &$\kappa b \phi  F_{\mu\nu}\widetilde F^{\mu\nu}$ & $2b^2/m_{\phi}^2\times (0,0,0,1)$ & $2(1-\xi)^2 b^2 (1-\mu^2)/m_{\phi}^2$ \\ \hline
tensor & $\kappa b \phi^{\mu\nu} \overline T_{\mu\nu}$ & $b^2/2m_\phi^2\times (1,0,0,1)$ & $(1-\xi)^2 b^2/2m_\phi^2$ \\ \hline
heterotic string & N/A & $\alpha'/16\times (3 , 1 , 0, 3)$ & $(\alpha'/16)(3\!+\!4\xi\!+\!13\xi^2)$
\end{tabular}
\caption{Higher-dimension operator coefficients $d_i$ and black hole entropy shift $\Delta S$ induced at low energies in various field theory UV completions defined by a massive particle $\phi$ that can be either a scalar, pseudoscalar, or tensor field, as well as the heterotic string.   In the former, the Lagrangian is ${\cal L}_{\rm UV} = \overline{\cal L} + {\cal L}_{\rm kin} + {\cal L}_{\rm int}$ where ${\cal L}_{\rm kin}$ is the canonical kinetic term for $\phi$.  The tensor field has a Fierz-Pauli mass term and couples to the energy-momentum tensor, $\overline T_{\mu\nu} = F_{\mu\rho}F_\nu^{\;\;\rho} - \frac{1}{4}g_{\mu\nu}F_{\rho\sigma}F^{\rho\sigma}$. All examples produce positive $\Delta S$ for any choice of parameters.
}\label{tab:UV}
\end{center}
\end{table*}

Furthermore, $d_3$ is bounded from below by $d_0$ and $d_9$. In particular, $\Delta S(\tfrac{1}{2},0,0)>0$ implies 
\be 
d_3/d_9>-1/30,
\ee
while $7\Delta S(\tfrac{1}{2},1,0)+5\Delta S(\tfrac{1}{2},-1,0) > 0$ implies 
\be 
d_3/d_0 > -1/30,
\ee 
which are the most stringent entropy bounds that can be placed on $d_3$ in terms of  either $d_0$ or $d_9$ alone.
A lower bound on $d_3$ is expected in light of the unitarity argument for positivity of $d_3$ given in \Ref{GB+}.

In addition, \Eq{eq:boundfull} implies that $d_6$, which vanishes for supersymmetric theories~\cite{CEMZ}, is bounded by  $d_0$ and $d_9$.  As illustrated in the last panel of \Fig{fig:boundspace}, regions outside of the allowed window in $d_6$ are excluded since there exist values of $\xi$ and $\mu$ such that $\Delta S(\xi,\mu,0)<0$ for any value of $d_3$.  We can obtain this bound analytically by taking $\Delta S(0,\mu,0)$ and marginalizing over $\mu$, yielding
\be
\left[\!\begin{array}{lr}
        2(d_9 \!-\! \sqrt{d_0 d_9}), & 4d_9 \leq d_0\\
       -d_0/2, & 4d_9 > d_0
        \end{array}\!\right]\!< d_6 < 2(d_9 \!+\! \sqrt{d_0 d_9}),
\ee
assuming both $d_0$ and $d_9$ are nonzero. If $d_0$ and $d_9$ are both generated at $\Lambda$, $|d_6|$ larger than $\sim 1/\Lambda^2$ is forbidden.
A similar conclusion, disallowing $d_6$ without the appearance of heavy states at the scale $|d_6|^{-1/2}$, was reached using causality in \Ref{CEMZ}.  

\section{UV Examples}
As a consistency check, we calculate the operator coefficients $d_i$ for various tree-level UV completions and verify that they obey the inequalities in \Eq{eq:boundfull}.  These include UV completions with a massive scalar or pseudoscalar coupled to the electromagnetic field strength and gravitational curvature tensor in all ways consistent with parity, as shown in \Tab{tab:UV}.
We also consider a UV completion with a massive tensor field with a ghost-free Fierz-Pauli mass term and minimal coupling to the energy-momentum tensor.  It is known that such a theory can have a cutoff parametrically higher than the mass of the tensor field~\cite{Hinterbichler:2011tt}.
In principle, other interactions with the tensor field could be possible, but such theories have not been well studied and we therefore have no reason to expect a consistent EFT, so such a generalization would not constitute a well defined check of our entropy bounds.  For example, a coupling such as $\phi^\mu_{\;\;\mu}F_{\nu\rho}F^{\nu\rho}$ would violate bounds coming from amplitude analyticity as well as entropy.
For a tree-level completion of \Eq{eq:DeltaL}, the massive state must be neutral, since charged lines cannot terminate in Feynman diagrams.

 As shown in \Tab{tab:UV}, the entropy shift is manifestly positive in our example scalar, pseudoscalar, and tensor completions,  providing  a check of our bounds.  In these examples, only $d_0$ and $d_9$ are generated.  
See the first panel of \Fig{fig:boundspace} for the parameter space spanned by these field theory UV completions.

Finally, we considered the low-energy effective theory of the heterotic string~\cite{Kats:2006xp,Gross:1986mw}, whose higher-dimension operators are not generated by quantum fields but nonetheless satisfy our bounds, as shown in \Tab{tab:UV}.  Here we have ignored the presence of the dilaton, but if this state is stabilized below the string scale, then its dynamics will be encoded in the massive scalar UV completion. 

\section{Weak Gravity Conjecture}
Our entropy bounds are intimately connected to the WGC. We define the extremality parameter 
\be 
\zeta =\sqrt{a^2+q^2+\widetilde q^2}/m = \sqrt{1-\xi^2}. 
\ee
In pure EM theory, black holes are described by the KN metric, which is free from naked singularities provided $\zeta \in [0,1]$.
The higher-dimension operators in \Eq{eq:DeltaL} correct the equations of motion, changing the allowed range of physical values to $\zeta \in[0,1+\Delta \zeta]$, as shown for the case of nonrotating electric black holes in \Ref{Kats:2006xp}.
As in \Ref{Cheung:2018cwt}, we calculate the extremality shift by writing $g^{rr}$ in terms of $\overline g^{rr}$ and the shifts in various parameters. Since $\partial_\theta \overline g^{rr} = \partial_\mu \overline g^{rr} = \partial_\nu \overline g^{rr}=0$ on the KN horizon and $\partial_r \overline g^{rr} = 0$ for the extremal horizon, we find that the extremality shift is 
\be 
\Delta \zeta = -\Delta g^{rr}/\partial_\zeta \overline g^{rr}|_{r_{\rm H}},
\ee 
evaluated on $r_{\rm H} = m(1+\xi)$. 
Similarly, for a black hole of fixed $(m,a,q,\widetilde q)$, the shift in the horizon is 
\be 
\Delta r_{\rm H} = -\Delta g^{rr}/\partial_r \overline g^{rr}|_{r_{\rm H}}.
\ee

In the extremal limit, the $\Delta r_{\rm H}$ term in the entropy shift dominates, since $\partial_r \overline g^{rr} \rightarrow 0$, so we have $\Delta S = (2\pi/\kappa^2) \Delta r_{\rm H} \partial_r \bar A$, where $\bar A=4\pi (r_{\rm H}^2 + a^2)$ is the area of the KN black hole. Using the fact that $\partial_\zeta \overline g^{rr}/\partial_r \overline g^{rr}|_{{\rm KN},r=r_{\rm H}} = m\zeta/\xi$, we directly relate the shift $\Delta \zeta$ in the extremality parameter to the entropy shift,
\be
\Delta \zeta(\mu,\nu) = + \frac{\kappa^2}{16\pi^2 m^2} \lim_{\xi \rightarrow 0}[\xi \Delta S(\xi,\mu,\nu)].\label{eq:Dz}
\ee
Sans rotation, we find that $\Delta \zeta \propto +d_0$ for the electric black hole \cite{Cheung:2018cwt}, $\Delta \zeta \propto +d_0 + 2d_6$ for the magnetic black hole, and $\Delta \zeta \propto +d_9$ for $q = \widetilde q$.

\begin{figure}[t]
\begin{center}
\includegraphics[width=3.8cm]{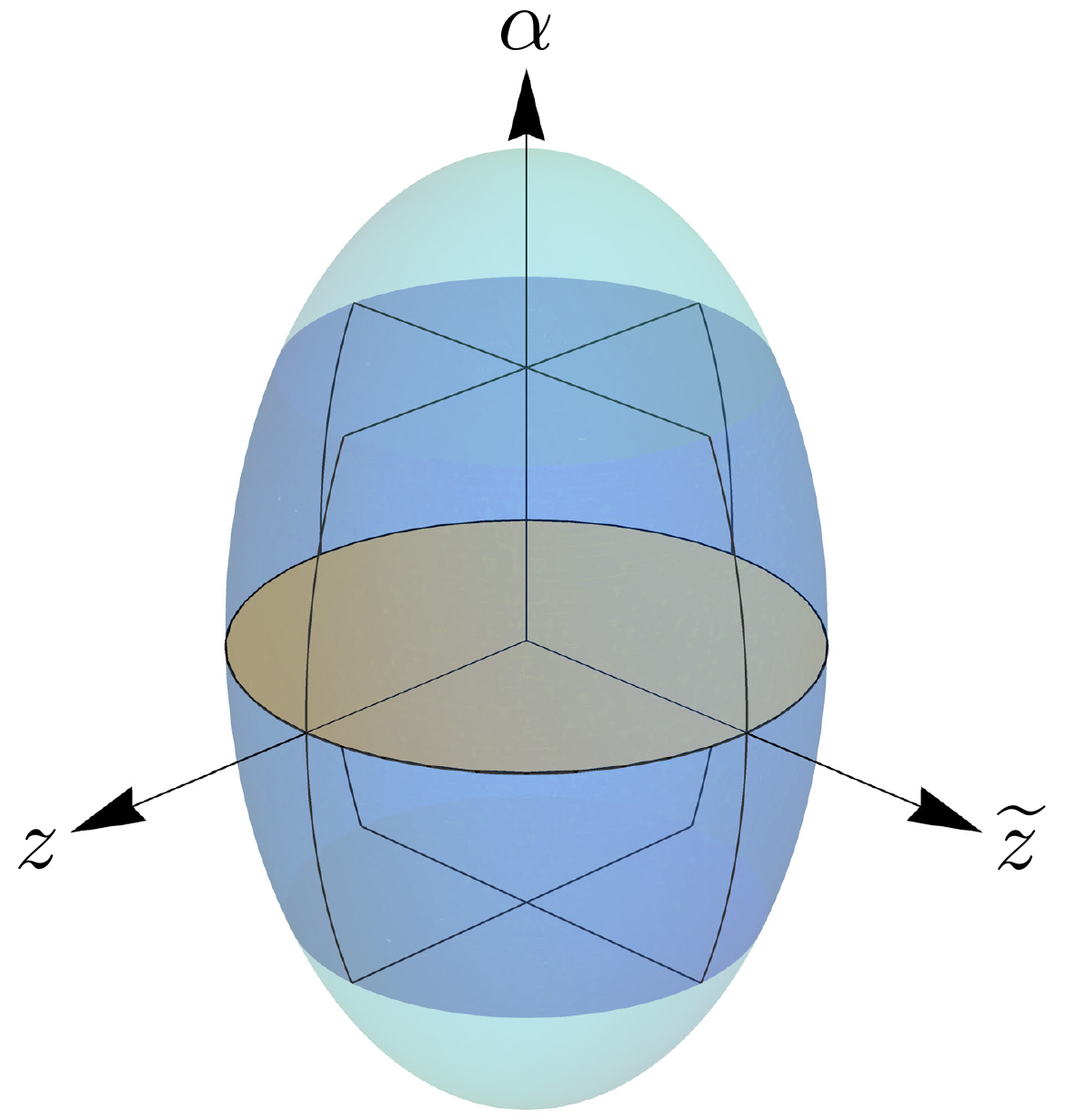}
\end{center}
\vspace{-6mm}
\caption{The cyan spheroid $z^2 + \widetilde z^2 + \alpha^2 \kappa^4/m^4 \leq 1$ denotes the space of all possible black holes without naked singularities.  The blue barrel-shaped region incorporates the additional criterion $|\alpha| < m^2/\sqrt{3}\kappa^2$, which defines the subspace of black holes that, due entropy bounds, can decay directly to lighter, extremality-corrected black holes. The gray disk represents all nonrotating dyonic black holes. }
\label{fig:ellipsoid}
\end{figure}

Let us now consider the decay of a black hole to lighter states with masses and charges $m_i,q_i,\widetilde q_i$ in natural units and angular momenta $j_i = 8\pi m_i a_i/\kappa^2$ including both intrinsic spin and orbital components.
Charge conservation implies that $\sum_i (q_i, \widetilde q_i) =(q, \widetilde q)$, while conservation of energy requires $\sum_i m_i < m$, where the inequality is needed if there is to be nonzero phase space for the decay products.  Angular momentum conservation requires $J = aM \leq \sum_i j_i$, where the inequality is due to the possibility of misaligned angular momenta.
Defining unitless ratios 
\be 
\begin{aligned}
z &= q/m\\
\widetilde z &= \widetilde q/m\\
 \alpha &= a m \kappa^{-2} = J/8\pi,
 \end{aligned}
 \ee
the set of KN black holes free of naked singularities forms a spheroid,
\be 
\zeta^2 = z^2 + \widetilde z^2 + (\alpha^2\kappa^4/m^4) \leq 1,
\ee
depicted in \Fig{fig:ellipsoid}.
Defining the analogous ratios for the decay products,
\be 
\begin{aligned}
(z_i,\widetilde z_i ) &= (q_i/m_i, \widetilde q_i/m_i) \\
\alpha_i &= a_i m \kappa^{-2} = j_i/8\pi \sigma_i,
\end{aligned}
\ee 
writing $\sigma_i = m_i/m$, we have 
\be 
\sum_i (\sigma_i z_i,\sigma_i \widetilde z_i) = (z ,\widetilde z) \text{  and  } \alpha \leq \sum_i \sigma_i \alpha_i.
\ee
Putting this all together to define $\mathbf w = (z, \widetilde z,\alpha)$, we conclude that a state with vector $\mathbf w$ must decay to states $\mathbf w_i$ for which $\mathbf w$ is either in the convex hull of the $\mathbf w_i$ or strictly between that convex hull and the $\alpha = 0$ plane.

The bound in \Eq{eq:boundfull} implies by \Eq{eq:Dz} that 
\be 
\Delta \zeta(\mu,\nu) > 0 \text{  for all  } \nu \in[0,1/\sqrt{3}).
\ee
Thus, extremal black holes located in the barrel-shaped region with $|a|/m = |\alpha|\kappa^2/m^2 <1/\sqrt{3}$ can decay directly to other rotating black holes.
The lighter rotating dyonic black holes form a closed surface everywhere outside the barrel-shaped region, so the tower of stable large black holes collapses to the scale at which the EFT breaks down.
In this sense, the WGC bounds implied by extremality corrections point to a principle of black hole self-sufficiency in providing their own decay channel.
While our $\Delta S >0$ bounds do not apply to black holes with $\nu>1/\sqrt{3}$, this is due to the fact that such states are known to be already thermodynamically unstable, even in the extremal limit~\cite{Monteiro:2009tc}.  In particular, these objects can shed their spin by emitting gravitons in states with nonzero angular momentum, decaying to the $\alpha = 0$ plane.
For nonrotating black holes, our entropy bound implies a dyonic version of the WGC: the set of near-extremal dyonic black holes can decay because they form a closed curve slightly outside the unit circle in the $\alpha = 0$ plane.

If the leading higher-dimension operators in \Eq{eq:DeltaL} have vanishing coefficients, the entropy will be corrected by yet-higher-derivative operators, which are still constrained by our reasoning since \Eq{eq:Dz} implies that $\Delta S \propto \Delta \zeta$ in the extremal limit~\cite{GrantNima}. While \Ref{Goon:2016mil} found $R^3$ operators generated either sign after integrating out massive matter with various spins at one loop, this does not contradict our results; these massive fields also induce four-derivative operators of the form in \Eq{eq:DeltaL}, which dominate over the $R^3$ operators within the regime of validity of the EFT. Moreover, the $\Delta S>0$ result has been strictly proven only for tree completions~\cite{Cheung:2018cwt}, though we conjecture it holds more generally.
If $\Delta S$ strictly vanishes to all orders, then there is a flat direction in the Euclidean action, indicating a symmetry protecting the states.  While such a symmetry is not present for EM theory black holes, it would be interesting to see how supersymmetry or massless moduli alter this picture.

\vspace{5mm}
\begin{acknowledgments}
We thank Nima Arkani-Hamed, John Preskill, Matt Reece, and Cumrun Vafa for useful discussions and comments. C.C. and J.L. are supported by the U.S. Department of Energy under award number DE-SC0011632. J.L. is also supported by the Institute for Quantum Information and Matter (IQIM) at Caltech, an NSF Physics Frontiers Center. G.N.R. is supported by the Miller Institute for Basic Research in Science at the University of California, Berkeley.
\end{acknowledgments}
\vspace{1cm}
\vfill
\newpage

\appendix

\section{Dyonic Metric Perturbation}\label{app}
In the nonrotating case, we can use a spherically-symmetric ansatz for the black hole metric,
\be
{\rm d}s^2 = -f(r) {\rm d}t^2 + {\rm d}r^2/g(r) + r^2 {\rm d}\Omega^2.
\ee
Following Refs.~\cite{Campanelli:1994sj,Kats:2006xp}, we can write $f(r)$ and $g(r)$ as integrals over the perturbed energy-momentum tensor $\overline T_{\mu\nu} + \Delta T_{\mu\nu}$ that appears on the right-hand side of the perturbed Einstein equation, $R_{\mu\nu} - \frac{1}{2}Rg_{\mu\nu} = \kappa^2(\overline T_{\mu\nu} + \Delta T_{\mu\nu})$.  
The shift $\Delta T_{\mu\nu}$ can be written as the sum of two contributions, one coming from the metric variation of $\Delta {\cal L}$ and one coming from the change to the field strength and metric in the original Maxwell stress tensor.

The shift in field strength is dictated by the perturbed Maxwell's equations, $\nabla_\nu F^{\mu\nu} = \nabla_\nu N^{\mu\nu}$, writing $N^{\mu\nu} = 2 \delta\Delta {\cal L}/\delta F_{\mu\nu}$.  Due to spherical symmetry, the only independent nonzero components of $F^{\mu\nu}$ are $F^{tr}$ and $F^{\theta\phi}$, so the Bianchi identity implies that $\nabla_r F^{\theta\phi} = 0$ in the perturbed solution, which means that $F^{\theta\phi}\propto 1/r^2$. Since $N^{\mu\nu}$ encodes the effects of higher-derivative terms, it cannot have the Coulomb scaling in $r$, so solving the perturbed Maxwell equation requires that $\nabla_\theta N^{\theta\phi} = 0$, which implies $N^{\theta\phi}\propto \csc \theta$. Thus, the magnetic component of the field strength is unchanged by the higher-derivative terms, $F^{\theta\phi} = \overline F^{\theta\phi}$ at ${\cal O}(d_i)$, while the electric component $F^{tr}$ is indeed modified.

We then jointly solve the modified Einstein and Maxwell equations at ${\cal O}(d_i)$:
\begin{widetext}
\be 
\begin{aligned}
f(r)&=1-\frac{2m}{r} + \frac{q^2+\widetilde q^2}{r^2} -\frac{4}{5r^6}\left\{
\begin{array}{l}
\phantom{+} (d_2+4d_3) (q^2 + \widetilde q^2)(q^2+\widetilde q^2 - 5mr+5r^2)\\
-10d_4 (q^2 - \widetilde q^2) (2q^2+2\widetilde q^2 -3mr+r^2) \\-d_5 [2(q^2+\widetilde q^2)(2q^2 - 3\widetilde q^2)-5 m r(q^2 - 2\widetilde q^2)-5 \widetilde q^2 r^2]\\
 +d_6(q^2 + \widetilde q^2)(q^2 + 3\widetilde q^2 - 5mr + 5r^2)
 \\ +4d_7(q^2 - \widetilde q^2)^2 + 2d_8(q^4 + \widetilde q^4)
\end{array} \right\}\\
g(r)&=1-\frac{2m}{r} + \frac{q^2+\widetilde q^2}{r^2} -\frac{4}{5r^6}\left\{
\begin{array}{l}
\phantom{+} (d_2+4d_3) (q^2 + \widetilde q^2)(6q^2+6\widetilde q^2 - 15mr+10r^2)\\
+10d_4 (q^2 - \widetilde q^2) (3q^2+3\widetilde q^2 -7mr+4r^2) \\+d_5 [(q^2 + \widetilde q^2)(11q^2 - 4 \widetilde q^2)-5 m r(5q^2 - 2\widetilde q^2)+5r^2(3q^2 - \widetilde q^2)]\\
 +d_6[2(q^2 + \widetilde q^2)(8q^2 - \widetilde q^2)-5mr(7q^2 - \widetilde q^2)+20q^2 r^2]
 \\ +4d_7(q^2 - \widetilde q^2)^2 +2d_8(q^4 + \widetilde q^4)
 \end{array} \right\}.
\end{aligned}
\ee
\end{widetext}

\bibliographystyle{utphys-modified}
\bibliography{dyonic}

\end{document}